\pdfoutput=1

\documentclass{article}

\usepackage{spconf,amsmath,graphicx}

\usepackage{epstopdf}

\usepackage{enumitem, textcomp, color, array}

\usepackage{ifpdf}

\usepackage{kotex}
\usepackage{xcolor}

\title{Enhancing music features by knowledge transfer \\
from user-item log data}

\name{Donmoon Lee$^{1,2,3}$, Jaejun Lee$^{1}$, Jeongsoo Park$^{2}$, and Kyogu Lee$^{1,3}$}
\address{$^{1}$Music and Audio Research Group, Seoul National University, Korea \\
$^{2}$Cochlear.ai, Korea\\
$^{3}$Center for Superintelligence, Seoul National University, Korea\\
}

\begin{document}
\ninept

\newcommand{\red}[1]{{\color{black}#1}}
\newcommand{\gnu}[1]{{\color{red}#1}}
% \newcommnad{\gnu}[1]{{\color{red}#1}}

%
\maketitle
\begin{abstract}
In this paper, we propose a novel method that exploits music listening log data for general-purpose music feature extraction. 
Despite the wealth of information available in the log data of user-item interactions, it has been mostly used for collaborative filtering to find similar items or users and \red {was not fully investigated for content-based music applications.}
We resolve this problem by extending intra-domain knowledge distillation to cross-domain: \textit{i.e.}, by transferring knowledge obtained from the user-item domain to the music content domain.
% The proposed system trains the model \red{to} estimate the log-based information from the audio contents and helps other task-specific models.
\red{The proposed system first trains the model that estimates log information from the audio contents; then it uses the model to improve other task-specific models.}
The experiments on various music classification and regression tasks show that the proposed method successfully improves the performances of the task-specific models.

\end{abstract}
\section{Introduction}
\label{sec:intro}

Recent advances in deep learning have shown to be very successful in many application areas, and music information retrieval (MIR) is not an exception. Remarkable performance gains seen in several MIR-related tasks \red {such as genre classification, mood estimation, automatic music transcription, and music recommendation}, to name a few, are mainly due to powerful deep learning models trained on a large amount of data. One of the major drawbacks in this approach is, however, that the performance of the deep learning-based algorithms is highly dependent on the amount of annotated training data.

The process of labeling and verifying music data requires more time and effort than other types of data, such as images. 
% This is mainly because playback of audio data takes time.
Furthermore, the fact that one music piece can have multiple attributes or labels make the process even more difficult. Some publicly available datasets such as GTZAN \cite{tzanetakis2002musical} contain manually labeled tags; hence the quality of the tags are guaranteed. However, the types of tags, as well as the amount of the data, are limited. The datasets such as million song dataset \cite{bertin2011million} and MagnaTagATune \cite{law2009evaluation} provide the user-generated label. They are known to provide a wide variety of unique tags; however, they can cause false negative problems since the users create the labels based on the \red{their own} criteria, often skipping some crucial information. Hence, the conventional research efforts were limited to use the tags that frequently appear \cite{choi2016automatic, lee2017multi}.

We aim to alleviate the above-mentioned problems of the audio data with the use of user listening log data (\textit{user-log}). This approach is mainly motivated by the previous music recommendation research which used audio content analysis to estimate missing user-log \cite{van2013deep} and the follow-up studies \cite{van2014transfer, lee2018deep}. According to Lee \textit{et al.} \cite{lee2018deep}, the user-log contains abundant useful information; their method based on the low-dimensional song representations derived from the user-log shows similar result compared to the audio-based approaches in the auto-tagging task. Besides, the main advantage of user-log is that unlike music tags, it can be collected without additional human efforts. It is constantly accumulating through the listening behavior of people.

\begin{figure}[htb]
  \centerline{\includegraphics[width=8.4cm]{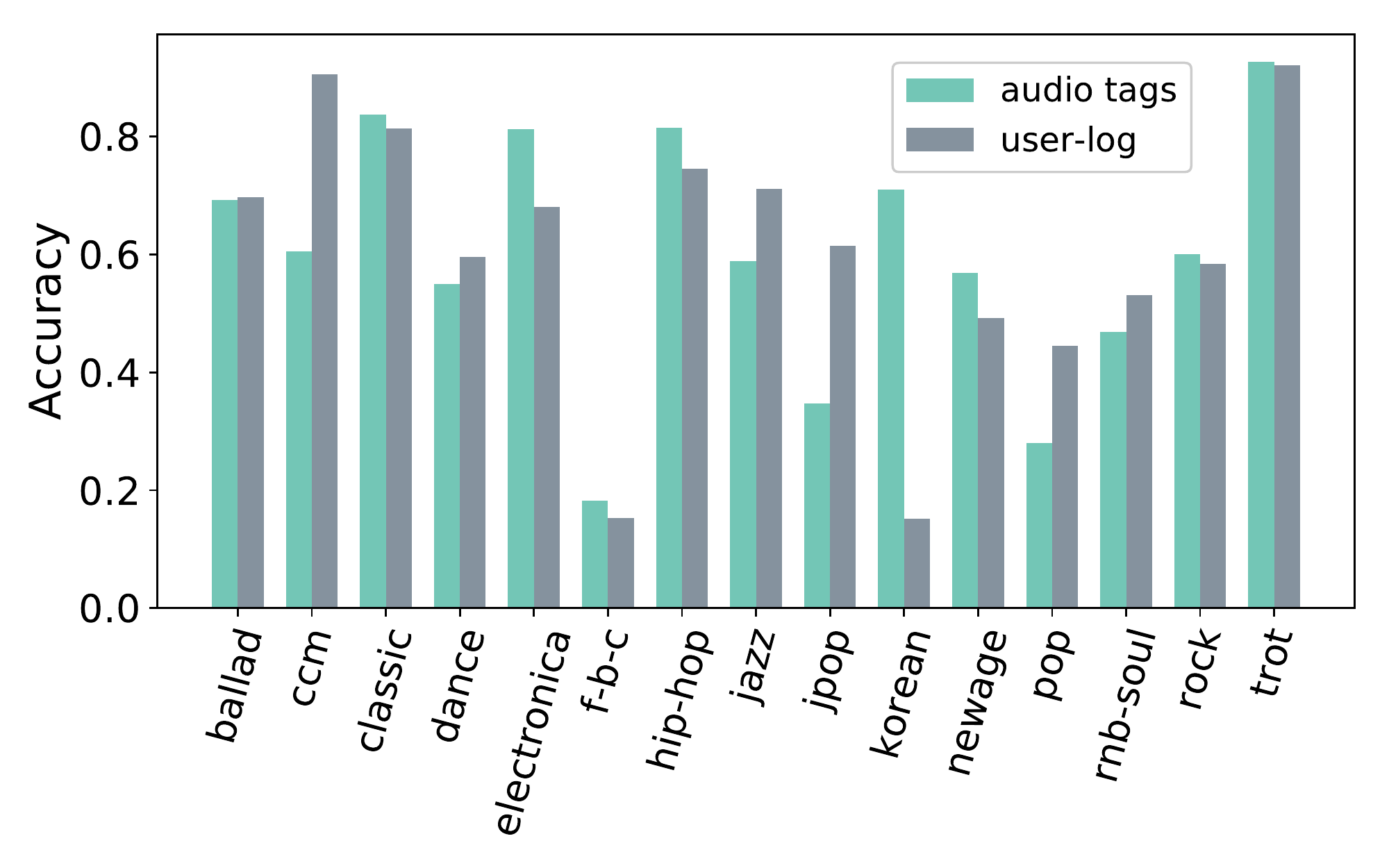}}
\caption{Class-wise results of genre classification using user-log and audio tags in Melon dataset. (korean and f-b-c indicate \textit{Korean traditional music} and the union genre of folk, blues, and country, respectively.)}
\label{fig:class_wise}
\end{figure}

% \red{Our pilot experiment shows that neural networks can learn different types of information from user-log compared to audio tags (Figure \ref{fig:class_wise}). We used two approaches to predict the genre in audio. The first is a conventional neural network, which uses genre tags directly as the ground-truth of the network. The second is a simple linear classifier using predicted user-log embedding trained through regression task. Both approaches have the same convolutional neural networks based feature extraction structure which is described in Section 2. Even if the second approach does not use the genre information in network training, it successfully predicts the genre and the class-wise performance differs from the case of the using audio tags.}
Figure \ref{fig:class_wise} shows our pilot experiment to test whether we can extract different types of information from the user-log data compared to the audio tags. The first model ('audio tags') is trained with the genre tags provided in the dataset; the second model ('user-log') is trained with the user-log embedding and concatenated to the linear classifier. Even if the second approach does not use the genre information in network training, it successfully predicts the genre and the class-wise performance differs from the case of using audio tags.
This suggests joint exploitation of the user-log data with the audio tags to be promising. 
Nevertheless, the previous approaches failed to simultaneously exploit the user-log data with the audio tags due to the difficulty in the cross-domain data conversion. 

We tackle this problem with \red{the training approach inspired by the} \textit{knowledge transfer}. 
Knowledge transfer in the neural network implies the use of the weight of the learned model for other tasks.
It is generally assumed that knowledge transfer models have extensive information because they are trained from large dataset or because the model capacity is large.
Depending on the methods to utilize a trained model, it can be classified as transfer methods and distillation methods.
The transfer methods directly use the outputs of the learned networks. Some methods exploited their intermediate activation as a feature for different tasks \cite{choi2017transfer, van2014transfer}; while others used their outputs for the fine-tuning \cite{gomezjazz}. On the other hand, the distillation methods, often referred to as teacher-student methods, utilize a posterior distribution
of the teacher network itself to train other models \cite{hinton2015distilling}.
Previous distillation-based approaches were proposed to solve within-domain applications \cite{romero2014fitnets, urban2016deep} or semi-supervised tasks \cite{papernot2016semi, meng2018adversarial} with the purpose of the network compression. 

In this paper, we introduce a novel method that transfers the knowledge from the user-log to music features.
To our best knowledge, this is the first approach to use the user-log data in the frame of knowledge \red{transfer.}
% distillation. 
We train a model to predict the user-log information from audio, transfer it to other task-specific models, and compare our approach with the conventional transfer methods for various music-related tasks.

\section{\red{Proposed Method}}
\label{sec:proposed}

\begin{figure}[tb]
  \centerline{\includegraphics[width=8.8cm]{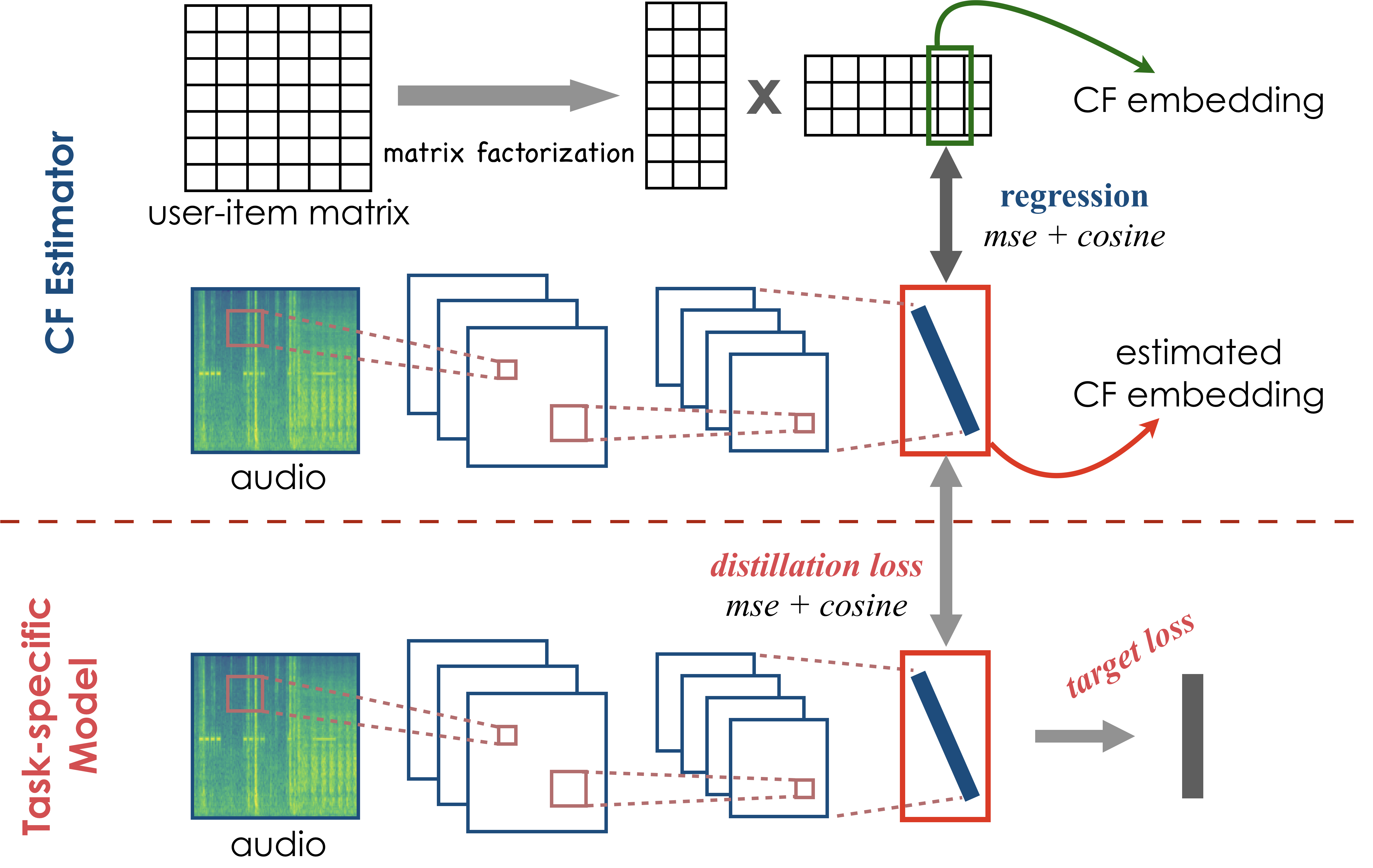}}
\caption{The overview of the proposed system. The CF estimator model is first trained to estimate user-log based representation for a given song. A task-specific model is then simultaneously trained to estimate both their objective and and CF embedding.}
\label{fig:proposed}
\end{figure}

Figure \ref{fig:proposed} shows an overview of the proposed method. 
It is designed to improve the performance of general music-related tasks by utilizing user-log information.

The user-log specifically exists as a user-item matrix indicating whether a particular user has listened to a particular song. 
Since the raw user-log is sparse and high-dimensional, it is usually converted into a compressed representation through dimension reduction methods such as matrix factorization. 
This is the typical representation of the song in the collaborative filtering (CF) method in the music recommendation systems, which is referred to as CF embedding.

One of the advantages of using CF embedding vectors is their size; it does not require human labor to generate the ground-truth. Therefore, it is relatively easy to obtain the CF embedding vectors from the large-scale music database compared to the other features that require human labeling. Nevertheless, it has only been used for the music recommendation systems since the method to utilize the CF embedding vectors has not been fully investigated. 

In order to successfully utilize the CF embedding vectors, we employ knowledge transfer method. To this end, we first train the model (CF estimator) that estimates the CF embedding vectors from the audio contents using massive user-log data; then we use the CF estimator's output to regularize the task-specific models which are trained with small-scale datasets. In specific, we constrain the penultimate layers of the task-specific models by using the estimated CF embedding to calculate additional loss function.

\section{Experiments}
\label{sec:exp}

% \subsection{Experiment Overview}
% We investigate the validity of the proposed method by applying it to the classification and the regression tasks.
% To this end, the CF estimator is trained with the CF embedding vectors correspond to the audio data in the large-scale music database.
% Then we
% First, a CF embedding is calculated by using large-scale music and log data, and the CF estimator model is trained to extract log-based information from audio content.
% The acquired information is applied to various music-related tasks to confirm the effect of the proposed system with the common transfer methods. 
% Additional experiments using different capacity models are conducted to see the effect of transfer methods on overfitting.

\subsection{Training CF estimator}

\subsubsection{Dataset}
Music and log data from Melon\footnote{A music streaming service in Korea, https://www.melon.com} are used to train the network.
The songs are randomly drawn from the Melon's preview excerpts which have played more than \red{once} in May 2018.
A single log contains information about who listened to which song. The number of the logs collected is about 4.7M, and when converting it to a binary user-item matrix, there are 5M users and 2.5M songs with non-zero values of 991M.
We optimized this using ALS algorithm based on the settings of the previous research \cite{hu2008collaborative}. The CF embedding vector dimension is set to 40 according to our experience on the music recommendation task.
% which is practically favorable for recommendation task.
We \red {randomly} selected 244,975 songs for the experiment, which was divided into training set (187,404 songs), validation set (20,831 songs), and test set (36,740 songs).

\subsubsection{Network architecture}
The structure of the teacher network is illustrated in Table \ref{tab:network}. It uses 30-second waveform as input, which is converted to mel spectrogram with a shape of (96, 1280, 1), each of which corresponds to the number of mel bins, time frames, and channels. 
The network consists of a repetitive stack of \texttt{double\_conv} blocks and max-pooling layers with the fully-connected output layer \cite{lee2017ensemble}. All the \texttt{double\_conv}
block consist of two sets of batch normalization, ReLU activation, and convolutional layer and one SE (Squeeze-and-excitation) \cite{hu2017squeeze} block following last convolutional layer.

All convolutional layer in one model has the fixed filter size of (3x3) and the same number of the feature maps. All the ratio of SE block is set to 8.
The size of the first four max-pooling layers are (4,5), (3,4), (2,4), and (2,4) in order. 
A global average pooling layer is attached to end of the fifth \texttt{double\_conv} block with following 40-dimensional fully-connected layer with linear activation.
The fully-connected layer is the output of the teacher network, and the dimension corresponds to that of the CF embedding.

As a loss function, additional cosine proximity was used in addition to the mean squared error used in previous studies \cite{van2013deep, lee2018deep}. Since the
direction is one of the important factors when dealing with the CF embedding.

We controlled the number of channels in the convolution layer to see the effect of model capacity following \cite{choi2017convolutional}. Four models with 32, 64, 128, and 256 channels were trained. 

\begin{table}[!tb]
\renewcommand{\arraystretch}{1.2}
\centering
\begin{tabular}{ >{\centering\arraybackslash}p{3.5cm} |  >{\centering\arraybackslash}p{3.2cm} }

    \hline
    \hline
    \bf{Layer}  & \bf{Output shape} \\
    \hline

    Audio input & (1,480000)\\
    Mel spectrogram & (96, 1280, 1) \\
    \texttt{Double\_conv} block & {} \\
    4 x 5 Max-pooling & (24, 256, \textit{F}) \\
    \texttt{Double\_conv} block & {} \\
    3 x 4 Max-pooling & (8, 64, \textit{F}) \\
    \texttt{Double\_conv} block & {} \\
    2 x 4 Max-pooling & (4, 16, \textit{F}) \\
    \texttt{Double\_conv} block & {} \\
    2 x 4 Max-pooling & (2, 4, \textit{F}) \\
    Global average pooling & (\textit{F}) \\
    Fully-connected layer & (40)\\

    \hline    
    \hline   

\end {tabular}
\caption{The structure of the CF estimator network. The network is constructed using the successive \texttt{double\_conv} block and max-pooling layer. \texttt{Double\_conv} block has two stacks of batch normalization layer, ReLU, and convolutional layer and SE block with the ratio of 8 after the last convolutional layer.
All the convolutional layers have (3 x 3) of filter size and the same number of channels. \textit{F} means the number of channels.}
\label{tab:network}
\end{table}

\subsubsection{Training details}

We selected the middle-30 second, which is similar to previous studies \cite{choi2016automatic, lee2017multi}, of the preview excerpt and resampled to 16,000 Hz.

Mel spectrogram is extracted on GPU using \textit{Kapre} \cite{choi2017kapre} with the 512 FFT point\red{s} (32ms) and hop size of 375 (23ms).
For training, Adam \cite{kingma2014adam} optimizer with a learning rate of 0.001 was used, and we choose the model which shows the least loss in the validation set. The experiment is implemented in \textit{Keras} \cite{chollet2015keras} with \textit{Tensorflow} \cite{tensorflow2015-whitepaper} backend. 

\subsection{Transfer to other tasks}

\subsubsection{List of tasks}

Since the CF embedding vector is a song-level descriptor, it is proper to apply the proposed method to the song-level classification/regression tasks. Six datasets are used in our experiments; five of them are related to music and the other one contains acoustic scene data. The last dataset is used to investigate the applicability of the proposed method even when it is applied for the general audio classification task. 

\begin{itemize}[leftmargin=*]
    \setlength\itemsep{0em}
    \item Genre classification tasks on the EmoMusic \cite{soleymani20131000},  GTZAN \cite{tzanetakis2002musical}, ExtendedBallRoom \cite{marchand2016extended}, FMA small \cite{defferrard2016fma}, and FMA medium datasets.
    \item Music emotion regression task on the EmoMusic dataset. Two independent models are trained to predict valence and arousal.
    \item Acoustic scene classification task on DCASE2016 \cite{mesaros2016tut} dataset.
\end{itemize}

\subsubsection{Experiment configuration}

The purpose of the experiment is to investigate the transition in performance of the task-specific models when the proposed knowledge transfer method is applied.
A total of four training methods including the proposed method are used. 
The baseline method (\textit{base}) trains the network from scratch. 
The first transfer method (\textit{fix}) uses CF estimator as the feature extractor and only the output layer is trained for the regression or classification task.
The second transfer method (\textit{init}) uses weights of the CF estimator to initialize the weights of the task-specific model. In this case, the entire weights of the model have to be trained.
The proposed method (\textit{kd}) only adds distillation loss to \textit{base}. The loss function is calculated with the activation of the penultimate layer and the output of the CF estimator; the sum of the mean squared error and the cosine proximity is used to calculate the loss function.

Except for GTZAN and BallRoomExtended datasets, the pre-defined training and test sets are used for evaluation.
The two datasets without pre-defined set are evaluated using stratified 10-fold cross-validation. We use the models that have the least task-specific loss in training for evaluation. In a dataset without validation set or where we perform cross-validation, the validation set is selected out of the train set.

\begin{figure}[tb]
  \centerline{\includegraphics[width=8.6cm]{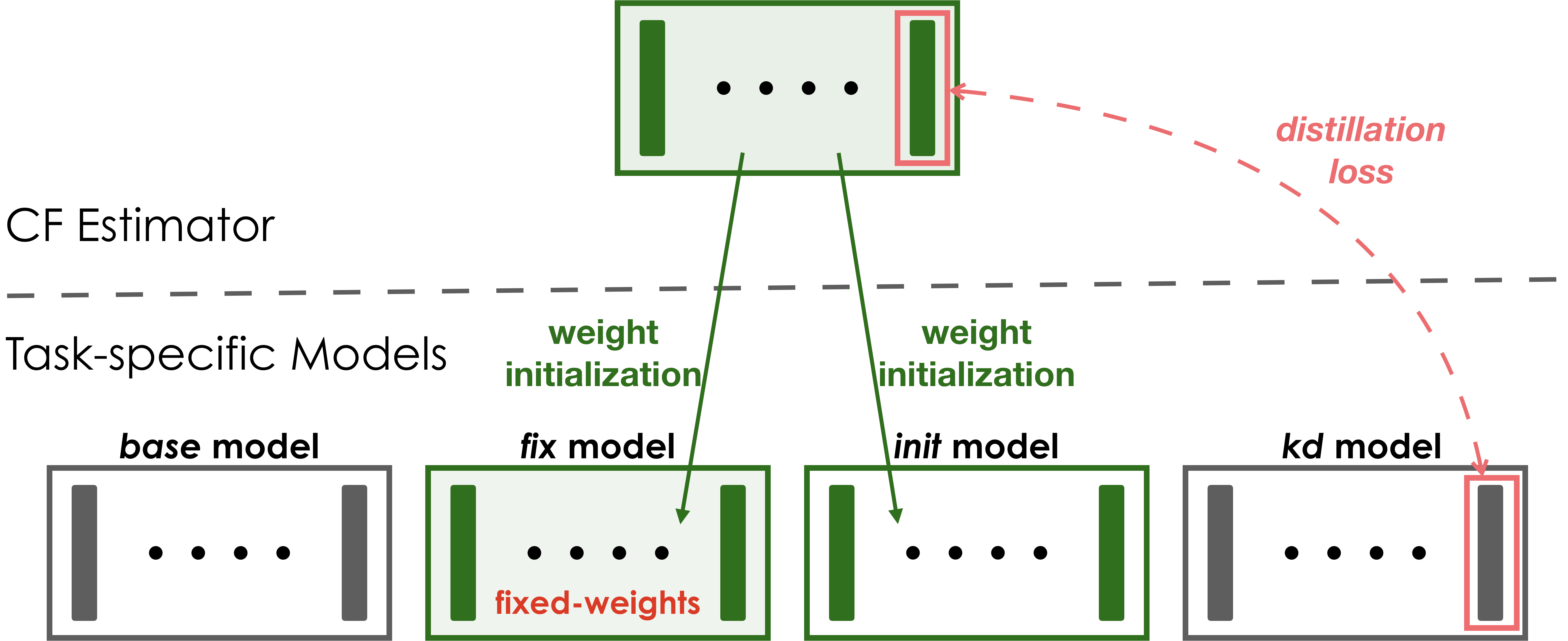}}
\caption{Four training \red{approaches} used in experiment. \textit{base} trains entire network from scratch, \textit{fix} only trains classifier, \textit{init} trains entire network with initialized weights, and \textit{kd} trains entire network with an additional loss function.}
\label{fig:exp}
\end{figure}

\begin{figure*}[htb]
  \centerline{\includegraphics[width=18cm]{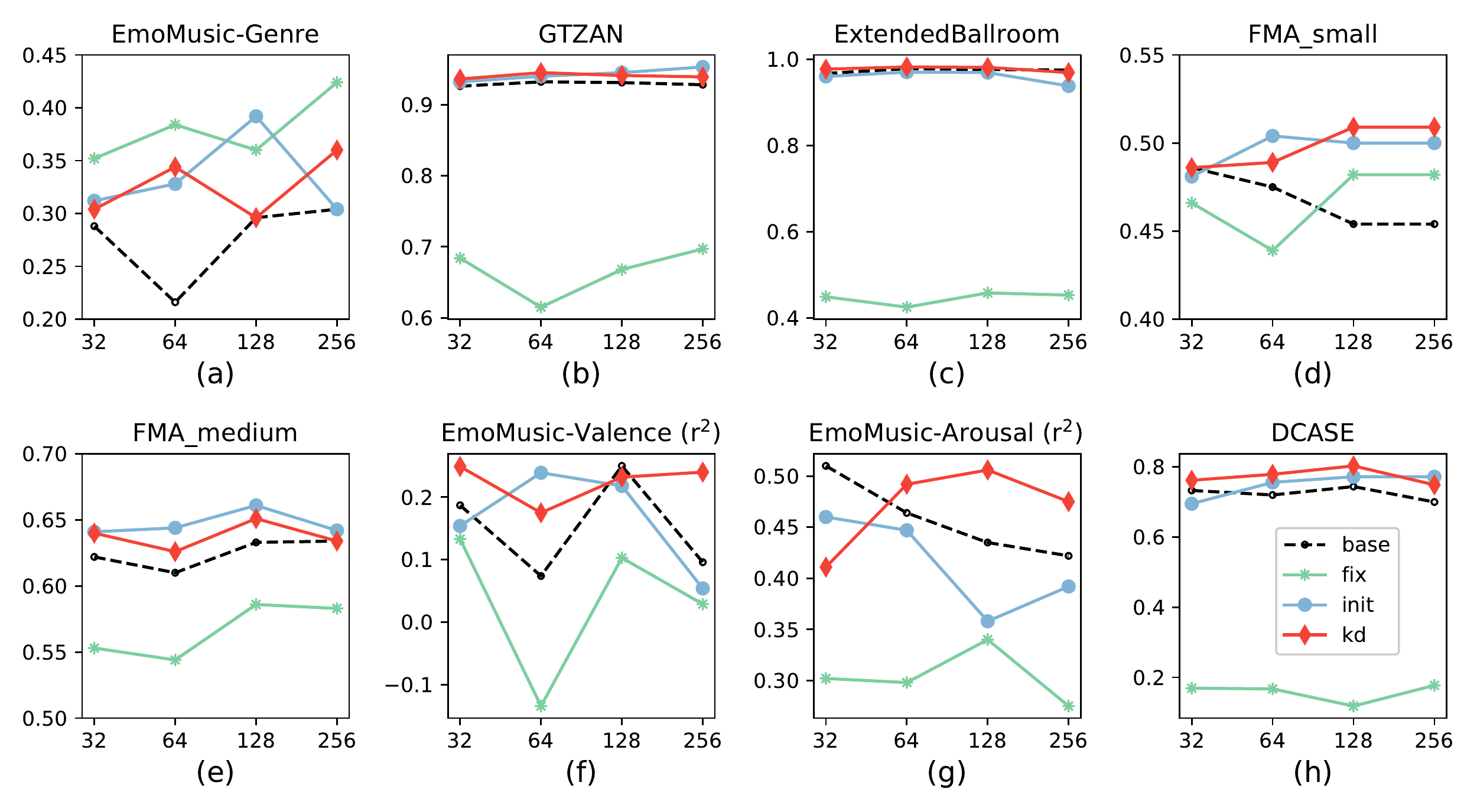}}
\caption{Knowledge transfer results on the various dataset. \textit{base} is condition without knowledge transfer, \textit{fix} direct uses the estimated CF embedding, \textit{init} initializes the weights using that of CF estimator, and \textit{kd} uses proposed distillation method. X-axis means the number of the channel, which is related to the model capacity. Y-axis means correlation coefficient ($r^2$) in mood detection tasks and the classification accuracy of the rest of tasks.}
\label{fig:res}
\end{figure*}

\section{Result and Discussion}
Experimental results are illustrated in the Figure \ref{fig:res}. Except the results in Figure \ref{fig:res} (b) and (c) which are evaluated through cross-validation, the remaining results are evaluated through single experiment.
That is, the results listed include variants of the experiment, and additional iterations are needed to determine statistical significance between methods.
However, we can see trends in each method through experimental results performed under various conditions.

In various datasets and experimental conditions, \textit{kd} shows not always, but mostly high performance. In particular, it always shows similar or higher performance than \textit{base}, except for the 32 channels of Figure \ref{fig:res} (g). 
For 24 experiments that performed classification tasks based on four model capacities in six datasets, \textit{kd} shows statistically significant improvement of 2.62 \% points over \textit{base} (Student paired t-test). 
\textit{init} also shows high performance in most experiments. However, the tendency of improvement is less stable than \textit{kd}, as there is a case where the performance is lower than \textit{base} as shown in Figure \ref{fig:res} (f). In fact, for all classification experiments, there is a performance improvement of 2.20 \% points over \textit{base}, which is slightly lower than that of \textit{kd}. However, there is no statistically significant difference between the two methods.
This suggests that the knowledge trained from user-log can be delivered through both the proposed method and the conventional transfer method.

The proposed method has an advantage in that it can be applied irrespective of the network structure. Because the distillation method is mainly used for the network compression, it is assumed that the structure of the task-specific network differs from that of the network which transfers the knowledge. Therefore, the proposed method is likely to be used in various ways in practical applications.
Another advantage is that it does not cause severe performance degradation. Even if the case of Figure \ref{fig:res} (h), where trained knowledge is not very helpful to the target task, the proposed method improves performance.
We assumed that the two loss functions competitively affect the task-specific model, and the network selectively receives useful information for the task.

CF embedding itself is not a generally well-defined descriptor for all music or audio related tasks. Since the direct use of the estimated CF embedding (\textit{fix}) always shows the lowest performance, except for the Figure \ref{fig:res} (a) and (d). In particular, the result of the acoustic scene classification task on Figure \ref{fig:res} (h), which is not a music-related task, shows slightly higher performance than a random guess.
Meanwhile, in Figure \ref{fig:res} (a), \textit{fix} shows the best accuracy in three out of four model capacities.
Considering EmoMusic's experimental results show low overall performance, it is assumed that the structure used in the experiment is not suitable for this data set.
We thought of the reason as the size of the dataset; EmoMusic is the smallest dataset in the experiment.
In this case, CF embedding rather than the direct genre label is better ground-truth for network training.
It suggests that estimated CF embedding contains the high-level information related to music and sometimes it is more powerful descriptor than the human labeled ground-truth. 

In a series of experiments, there is no noticeable trend between the capacity of the model and the transfer method.
We thought that model capacity would not have had a significant impact because regularization modules such as s batch normalization and SE blocks are used from \textit{base} structure. 
In fact, the results of Figure \ref{fig:res} (b) and (c) are similar to state-of-the-art regardless of the capacity of the models.
% Rather, most \textit{fix} have a similar trend to \textit{init} and \textit{kd}, especially \textit{kd}. 
Rather, there appears to be a relationship between learned knowledge and transfer effect.
There is a similar trend between \textit{fix} and \textit{init}, \textit{fix} and \textit{kd}, which is more prominent in the latter.
It implies that the type of learned knowledge has a great influence on knowledge transfer, especially the proposed method, but further studies are required to clarify the key variables associated with it.

\section{Conclusion}
\label{sec:conclusion}
In this study, we proposed the knowledge transfer method that uses user-log information for music feature extraction. We made song descriptors from the user-item log and trained the model that aims to predict the song descriptors from the audio contents. Then we tried to apply the learned knowledge to various task-specific networks. According to the experimental results, the proposed distillation method successfully transferred the information of the user-log to the task-specific network as it utilized the information as an additional loss. Our method has advantages in that it can be similarly applied to various music-related tasks as well as it is the first approach that successfully utilized user-log data in conjunction with the music content data.

\section{Acknowledgements}
The authors are deeply grateful to Keunwoo Choi for the help on paper writing. 
This work was supported partly by Kakao and Kakao Brain corporations and partly by Next-Generation Information Computing Development Program through the National Research Foundation of Korea (NRF) funded by the Ministry of Science and ICT (NRF-2017M3C4A7078548).

\vfill\pagebreak

% \begin{figure}[htb]

% \begin{minipage}[b]{1.0\linewidth}
%   \centering
%   \centerline{\includegraphics[width=8.5cm]{image1}}
% %  \vspace{2.0cm}
%   \centerline{(a) Result 1}\medskip
% \end{minipage}
% %
% \begin{minipage}[b]{.48\linewidth}
%   \centering
%   \centerline{\includegraphics[width=4.0cm]{image3}}
% %  \vspace{1.5cm}
%   \centerline{(b) Results 3}\medskip
% \end{minipage}
% \hfill
% \begin{minipage}[b]{0.48\linewidth}
%   \centering
%   \centerline{\includegraphics[width=4.0cm]{image4}}
% %  \vspace{1.5cm}
%   \centerline{(c) Result 4}\medskip
% \end{minipage}
% %
% \caption{Example of placing a figure with experimental results.}
% \label{fig:res}
% %
% \end{figure}

\bibliographystyle{IEEEbib}
\bibliography{refs}

\begin{thebibliography}{10}

\bibitem{tzanetakis2002musical}
George Tzanetakis and Perry Cook,
\newblock ``Musical genre classification of audio signals,''
\newblock {\em IEEE Transactions on speech and audio processing}, vol. 10, no.
  5, pp. 293--302, 2002.

\bibitem{bertin2011million}
Thierry Bertin-Mahieux, Daniel~PW Ellis, Brian Whitman, and Paul Lamere,
\newblock ``The million song dataset.,''
\newblock in {\em Ismir}, 2011, vol.~2, p.~10.

\bibitem{law2009evaluation}
Edith Law, Kris West, Michael~I Mandel, Mert Bay, and J~Stephen Downie,
\newblock ``Evaluation of algorithms using games: The case of music tagging.,''
\newblock in {\em ISMIR}, 2009, pp. 387--392.

\bibitem{choi2016automatic}
Keunwoo Choi, Gy{\"o}rgy Fazekas, and Mark Sandler,
\newblock ``Automatic tagging using deep convolutional neural networks,''
\newblock in {\em The 17th International Society of Music Information Retrieval
  Conference, New York, USA}. International Society of Music Information
  Retrieval, 2016.

\bibitem{lee2017multi}
Jongpil Lee and Juhan Nam,
\newblock ``Multi-level and multi-scale feature aggregation using pretrained
  convolutional neural networks for music auto-tagging,''
\newblock {\em IEEE signal processing letters}, vol. 24, no. 8, pp. 1208--1212,
  2017.

\bibitem{van2013deep}
Aaron Van~den Oord, Sander Dieleman, and Benjamin Schrauwen,
\newblock ``Deep content-based music recommendation,''
\newblock in {\em Advances in neural information processing systems}, 2013, pp.
  2643--2651.

\bibitem{van2014transfer}
A{\"a}ron Van Den~Oord, Sander Dieleman, and Benjamin Schrauwen,
\newblock ``Transfer learning by supervised pre-training for audio-based music
  classification,''
\newblock in {\em Conference of the International Society for Music Information
  Retrieval (ISMIR 2014)}, 2014.

\bibitem{lee2018deep}
Jongpil Lee, Kyungyun Lee, Jiyoung Park, Jangyeon Park, and Juhan Nam,
\newblock ``Deep content-user embedding model for music recommendation,''
\newblock {\em arXiv preprint arXiv:1807.06786}, 2018.

\bibitem{choi2017transfer}
Keunwoo Choi, Gy{\"o}rgy Fazekas, Mark Sandler, and Kyunghyun Cho,
\newblock ``Transfer learning for music classification and regression tasks,''
\newblock {\em arXiv preprint arXiv:1703.09179}, 2017.

\bibitem{gomezjazz}
Juan~S G{\'o}mez, Jakob Abe{\ss}er, and Estefan{\i}a Cano,
\newblock ``Jazz solo instrument classification with convolutional neural
  networks, source separation, and transfer learning,''
\newblock .

\bibitem{hinton2015distilling}
Geoffrey Hinton, Oriol Vinyals, and Jeff Dean,
\newblock ``Distilling the knowledge in a neural network,''
\newblock {\em arXiv preprint arXiv:1503.02531}, 2015.

\bibitem{romero2014fitnets}
Adriana Romero, Nicolas Ballas, Samira~Ebrahimi Kahou, Antoine Chassang, Carlo
  Gatta, and Yoshua Bengio,
\newblock ``Fitnets: Hints for thin deep nets,''
\newblock {\em arXiv preprint arXiv:1412.6550}, 2014.

\bibitem{urban2016deep}
Gregor Urban, Krzysztof~J Geras, Samira~Ebrahimi Kahou, Ozlem Aslan, Shengjie
  Wang, Rich Caruana, Abdelrahman Mohamed, Matthai Philipose, and Matt
  Richardson,
\newblock ``Do deep convolutional nets really need to be deep and
  convolutional?,''
\newblock {\em arXiv preprint arXiv:1603.05691}, 2016.

\bibitem{papernot2016semi}
Nicolas Papernot, Mart{\'\i}n Abadi, Ulfar Erlingsson, Ian Goodfellow, and
  Kunal Talwar,
\newblock ``Semi-supervised knowledge transfer for deep learning from private
  training data,''
\newblock {\em arXiv preprint arXiv:1610.05755}, 2016.

\bibitem{meng2018adversarial}
Zhong Meng, Jinyu Li, Yifan Gong, and Biing-Hwang Juang,
\newblock ``Adversarial teacher-student learning for unsupervised domain
  adaptation,''
\newblock in {\em 2018 IEEE International Conference on Acoustics, Speech and
  Signal Processing (ICASSP)}. IEEE, 2018, pp. 5949--5953.

\bibitem{hu2008collaborative}
Yifan Hu, Yehuda Koren, and Chris Volinsky,
\newblock ``Collaborative filtering for implicit feedback datasets,''
\newblock in {\em Data Mining, 2008. ICDM'08. Eighth IEEE International
  Conference on}. Ieee, 2008, pp. 263--272.

\bibitem{lee2017ensemble}
Donmoon Lee, Subin Lee, Yoonchang Han, and Kyogu Lee,
\newblock ``Ensemble of convolutional neural networks for weakly-supervised
  sound event detection using multiple scale input,''
\newblock {\em Detection and Classification of Acoustic Scenes and Events
  (DCASE)}, 2017.

\bibitem{hu2017squeeze}
Jie Hu, Li~Shen, and Gang Sun,
\newblock ``Squeeze-and-excitation networks,''
\newblock .

\bibitem{choi2017convolutional}
Keunwoo Choi, Gyorgy Fazekas, Mark Sandler, and Kyunghyun Cho,
\newblock ``Convolutional recurrent neural networks for music classification,''
\newblock in {\em 2017 IEEE International Conference on Acoustics, Speech, and
  Signal Processing, ICASSP 2017}. Institute of Electrical and Electronics
  Engineers Inc., 2017.

\bibitem{choi2017kapre}
Keunwoo Choi, Deokjin Joo, and Juho Kim,
\newblock ``Kapre: On-gpu audio preprocessing layers for a quick implementation
  of deep neural network models with keras,''
\newblock in {\em Machine Learning for Music Discovery Workshop at 34th
  International Conference on Machine Learning}. ICML, 2017.

\bibitem{kingma2014adam}
Diederik~P Kingma and Jimmy Ba,
\newblock ``Adam: A method for stochastic optimization,''
\newblock {\em arXiv preprint arXiv:1412.6980}, 2014.

\bibitem{chollet2015keras}
Fran\c{c}ois Chollet et~al.,
\newblock ``Keras,'' \url{https://keras.io}, 2015.

\bibitem{tensorflow2015-whitepaper}
Mart\'{\i}n Abadi, Ashish Agarwal, Paul Barham, Eugene Brevdo, Zhifeng Chen,
  Craig Citro, Greg~S. Corrado, Andy Davis, Jeffrey Dean, Matthieu Devin,
  Sanjay Ghemawat, Ian Goodfellow, Andrew Harp, Geoffrey Irving, Michael Isard,
  Yangqing Jia, Rafal Jozefowicz, Lukasz Kaiser, Manjunath Kudlur, Josh
  Levenberg, Dandelion Man\'{e}, Rajat Monga, Sherry Moore, Derek Murray, Chris
  Olah, Mike Schuster, Jonathon Shlens, Benoit Steiner, Ilya Sutskever, Kunal
  Talwar, Paul Tucker, Vincent Vanhoucke, Vijay Vasudevan, Fernanda Vi\'{e}gas,
  Oriol Vinyals, Pete Warden, Martin Wattenberg, Martin Wicke, Yuan Yu, and
  Xiaoqiang Zheng,
\newblock ``{TensorFlow}: Large-scale machine learning on heterogeneous
  systems,'' 2015,
\newblock Software available from tensorflow.org.

\bibitem{soleymani20131000}
Mohammad Soleymani, Micheal~N Caro, Erik~M Schmidt, Cheng-Ya Sha, and Yi-Hsuan
  Yang,
\newblock ``1000 songs for emotional analysis of music,''
\newblock in {\em Proceedings of the 2nd ACM international workshop on
  Crowdsourcing for multimedia}. ACM, 2013, pp. 1--6.

\bibitem{marchand2016extended}
Ugo Marchand and Geoffroy Peeters,
\newblock ``The extended ballroom dataset,''
\newblock 2016.

\bibitem{defferrard2016fma}
Micha{\"e}l Defferrard, Kirell Benzi, Pierre Vandergheynst, and Xavier Bresson,
\newblock ``Fma: a dataset for music analysis,''
\newblock {\em arXiv preprint arXiv:1612.01840}, 2016.

\bibitem{mesaros2016tut}
Annamaria Mesaros, Toni Heittola, and Tuomas Virtanen,
\newblock ``Tut database for acoustic scene classification and sound event
  detection,''
\newblock in {\em Signal Processing Conference (EUSIPCO), 2016 24th European}.
  IEEE, 2016, pp. 1128--1132.

\end{thebibliography}

\end{document}